\documentclass[a4paper]{jpconf}
\usepackage{graphicx,amsmath,color}

\usepackage{cite}

\newcommand{\ii}{\mathrm{i}}

\begin{document}

\title{Non-equilibrium current and electron pumping in nanostructures}

\author{A Alvermann and H Fehske}

\address{Institute of Physics, Ernst-Moritz-Arndt University, 17487 Greifswald, Germany}

\ead{alvermann@physik.uni-greifswald.de}

\begin{abstract}
We discuss a numerical method to study electron transport in mesoscopic devices out of equilibrium.
The method is based on the solution of operator equations of motion, using efficient Chebyshev time propagation techniques.
Its peculiar feature is the propagation of operators backwards in time.
In this way the resource consumption scales linearly with the number of states used to represent the system.
This allows us to calculate the current for non-interacting electrons in large one-, two- and three-dimensional lead-device configurations with time-dependent voltages or potentials. 
We discuss the technical aspects of the method and present results for an electron pump device and a disordered system, where we find transient behaviour that exists for a very long time and may be accessible to experiments.

\end{abstract}

Electron transport through mesoscopic devices contacted to leads
is intensely studied in chemistry and physics (see e.g. Refs.~\cite{Cuni05,Da95_2}).
The conceptual basis for theoretical studies is the non-equilibrium Green function (Keldysh) formalism~\cite{HJ08}. 
The Meir-Wingreen-formula~\cite{MW92} allows for the calculation of steady state currents.
For time-dependent potentials or gate voltages the calculation of a current is a serious problem already for non-interacting electrons.
One crucial point in most approaches is that the resource consumption scales quadratically with the number of system sites, since each Green function $G_{ij}(t,t')$ or operator product $c^\dagger_i(t) c^{}_j(t')$ has two site indices.
The necessity to study large systems conflicts with the rapid growth of computational demands, especially if long leads with long recurrence time are required.
Recent studies therefore addressed mainly one-dimensional (1D) situations,
e.g. by time-propagation of operator expectation values or of several single-particle eigenstates~\cite{DS03,AS07,SKRG08}.

In the present contribution we numerically solve the equation of motion for a single fermion operator.
In contrast to previous studies, the initial time coordinate is propagated backwards in time.
Then the computational effort, in particular the memory consumption, scales linearly with the number of lattice sites used to represent device and leads.
This allows us to calculate the non-equilibrium current even for large systems that could not be treated otherwise.

\begin{figure}
\centering
\includegraphics[width=0.5\textwidth]{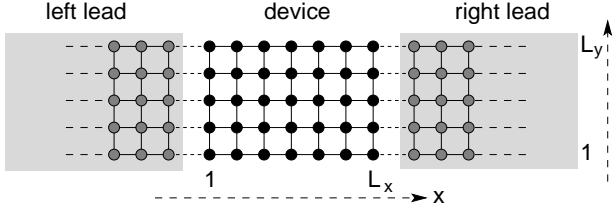}
\hspace*{2pc}
\begin{minipage}[b]{0.43\textwidth}
\caption{\label{slab}Typical 2D slab geometry, with a central device region contacted to a left/right lead.}
\end{minipage}
\end{figure}

We consider the following situation:
A device of $L_x \times L_y  \times L_z$-sites is contacted 
to two long leads extending along the $x$-direction (see figure~\ref{slab}).
This slab geometry includes the 2D ($L_z=1$) and 1D ($L_y=L_z=1$) case.
For the kinetic energy in the Hamiltonian
\begin{equation}
 H(t) = - \tilde{t} \sum_{\langle \mathbf{m}, \mathbf{n} \rangle} c^\dagger_\mathbf{m} c^{}_\mathbf{n}  +
\frac{U(t)}{2} \sum_{\substack{\mathbf{m} \\ x < 1}} 	c^\dagger_\mathbf{m} c^{}_\mathbf{m} \;
 + \sum_{\substack{\mathbf{m} \\ 1 \le x \le L_x}} V_\mathbf{m}(t) c^\dagger_\mathbf{m} c^{}_\mathbf{m} \;
- \frac{U(t)}{2} \sum_{\substack{\mathbf{m} \\ x > L_x}} c^\dagger_\mathbf{m} c^{}_\mathbf{m}
\end{equation}
we assume nearest-neighbour hopping  ($\propto \tilde{t}$) along each axis $i=x,y,z$, and set $\tilde{t}=1$ as the unit of energy.
The potential energy terms describe the voltage bias $U(t)$ applied between the left/right lead, and the local potentials $V_\mathbf{m}(t)$ in the device.
The current along a bond in the $x$-direction is given by the expectation value of the corresponding operator
\begin{equation}\label{jop}
\hat{\jmath}^x_{\mathbf{m}} = \ii \tilde{t} (c^\dagger_{\mathbf{m}+\mathbf{e}_x} c^{}_{\mathbf{m}} - c^\dagger_{\mathbf{m}} c^{}_{\mathbf{m}+\mathbf{e}_x}) \;.
\end{equation}

To specify the initial conditions, 
we assume that for times $t<0$ the device is isolated and uncharged, and the leads are in equilibrium at zero temperature (this situation can be described by setting $\tilde{t}=0$ along the contacts at $x=1,L_x$ and letting $V_\mathbf{m} \to \infty$ in $H(t<0)$).
At time $t=0$ device and leads are brought into contact,
and electrons can flow onto the device.
We then ask for the current $I(t)=\sum_{yz} \langle \psi(t)| \hat{\jmath}^x_{(x,y,z)} |\psi(t) \rangle$ through a section $x=const.$ of the slab.

Time propagation of the many-fermion Fock state $|\psi(t)\rangle$ is not feasible,
and the reduction of this problem to the propagation of several single-fermion states
is restricted to non-interacting electrons and introduces additional errors that must be controlled. 
It appears to be more natural to propagate the current operators themselves.
With the time evolution operator $\hat{U}(t,t')$, defined by
$\ii \frac{\partial}{\partial t} \hat{U}(t,t') = H(t) \hat{U}(t,t')$, $\hat{U}(t,t) = 1$,
operators can be transformed to the Heisenberg picture
\begin{equation}
  A^H(t,t_0) = \hat{U}(t_0,t) \, A \, \hat{U}(t,t_0) \;,
\end{equation}
and expectation values are obtained from $A^H(t,0)$ and the initial state $|\psi_0\rangle = |\psi(t=0)\rangle$ as
\begin{equation}\label{ExpectA}
 \langle \psi(t)| A | \psi(t) \rangle = \langle \psi_0 | A^H(t,0) |\psi_0 \rangle \;.
\end{equation}
Therefore we must represent $A^H(t,0)$ in terms of operators $A_j=A^H_j(0,0)$ at time $t=0$,
\begin{equation}\label{AExpansion}
  A^H(t,0) = \sum_j a(j,t,0)  A_j \;,
\end{equation}
which allows evaluation of equation~\eqref{ExpectA} since all initial expectation values 
$\langle\psi_0| A_j|\psi_0\rangle$ are known.

We observe that two ways exist to obtain $A^H(t,0)$.
The standard way starts from 
\begin{equation}\label{AFirstWay}
 \ii \frac{\partial}{\partial t} A^H(t,0) = \hat{U}(0,t) \, [A,H(t)] \, \hat{U}(t,0) = [A,H(t)]^H(t,0) \;.
\end{equation}
In this equation, the commutator is given as a combination of operators $A_j$, which is evaluated at time $t$ in the Heisenberg picture. Therefore solution of equation~\eqref{AFirstWay} requires time propagation of all operators $A^H_j(t,0)$. The total number of coefficients in the corresponding expansions equation~\eqref{AExpansion} for these operators grows quadratically with the number of sites.

To avoid quadratic growth we here proceed the opposite way, starting from
$- \ii \frac{\partial}{\partial t'} \hat{U}(t,t') = \hat{U}(t,t') H(t')$.
In contrast to the previous case, we do not need to transform the commutator in
\begin{equation}\label{Aback}
 \ii \frac{\partial}{\partial t'} A^H(t,t') = [H(t'),\hat{U}(t',t) A \hat{U}(t,t')] = [H(t'), A^H(t,t')] 
\end{equation}
to the Heisenberg picture. Instead, it is only the single operator $A$ whose expansion
\begin{equation}\label{AExpansion2}
  A^H(t,t') = \sum_j a(j,t,t')  A_j
\end{equation}
\newpage\noindent
must be propagated in time. We do not need to keep track of all the other operators $A^H_j(t,t')$. 
Therefore memory consumption and computational effort are proportional to the number of sites.
Making use of $(AB)^H(t,t')=A^H(t,t') B^H(t,t')$, each fermion operator in equation~\eqref{jop} can be propagated separately to obtain $(\hat{\jmath}^x_{\mathbf{m}})^H(t,t')$.

The price to be paid here is that equation~\eqref{Aback} leads to propagation backwards in time, evolving the second time coordinate from the initial $A(t,t)=A(0,0)=A$ to the final $A(t,0)$.
The main reason why we nevertheless consider this procedure is that, avoiding quadratic growth of memory consumption, only this way allows for the treatment of large systems, which are inaccessible in the first way.
With modern Chebyshev techniques~\cite{TK84}, time propagation itself is fast and accurate, and the additional effort for backward propagation is compensated by the reduction of the problem size.  
Note that for periodic time dependence, operators can be propagated iteratively by a full period.

Turning to the slab geometry, the eigenstates of decoupled infinite leads are discrete along the $y$-, $z$-direction and continuous along the $x$-direction.
In the actual calculation, infinite leads are replaced by long leads of length $L \gg L_{x,y,z}$.
The recurrence time revealing the artificial discretization is of the order $L$,
and must be larger than the maximal time $t$ in the calculations.
Using operators $c_\alpha$ for lead eigenstates to energy $\epsilon_\alpha$
and operators $c_\mathbf{m}$ for device sites $\mathbf{m}$ as the $A_j$ in equation~\eqref{AExpansion2}
the initial conditions specify the expectation values at time $t=0$,
\begin{equation}
  \langle c^\dagger_\mathbf{m} c^{}_\mathbf{n} \rangle = 
  \langle c^\dagger_\mathbf{m} c^{}_\alpha \rangle =0 \;, \quad
 \langle c^\dagger_\alpha c^{}_{\alpha'} \rangle = \delta_{\alpha \alpha'} \Theta(\mu-\epsilon) \;, 
\end{equation}
where $\mu$ is the Fermi energy.
Note that only these conditions change for finite temperatures.

\begin{figure}
\centering
\includegraphics[width=0.5\textwidth]{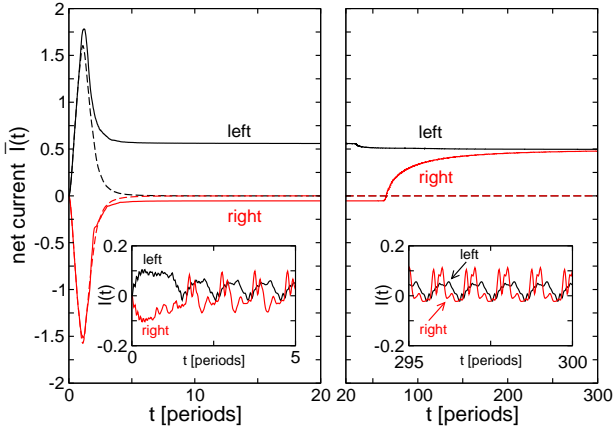}
\hspace*{2pc}
\begin{minipage}[b]{0.43\textwidth}
\caption{\label{pumping} (Colour online) Current $I(t)$ measured 10 sites to the left (black curve) or right (red curve) of a 1D pumping device with $L_x=500$ sites.
The system parameters are $V=2$, $\omega=\pi/10$, $k=\pi/2$, with Fermi momentum $k_F= \pi/4$ for the leads. Time $t$ is given in multiples of the period $T=2\pi/\omega=20$.
Shown is the net current $\bar{I}(t) = \int^t_{t-T} I(t') dt'$  over one period, in comparison to the situation with a static potential wave (dashed curves). The insets display $I(t)$ over the first and last 5 periods.
}
\end{minipage}
\end{figure}

In our first example in figure~\ref{pumping}, a travelling potential wave
\begin{equation}
  V_x(t) = V \cos(k x - \omega t)
\end{equation}
pumps electrons in a 1D system with zero voltage bias $U=0$ from the left to right (i.e. along the positive $x$-direction).
In the beginning electrons flow onto the initially uncharged device.
The initial transient behaviour evolves into a `pseudo' steady state (left panel), which persists over $\approx 50$ periods. The true steady state, with equal current at both sites of the device, is approached only in the extreme long time limit (right panel).
Comparison to the case of a static potential, with the same $k$ but $\omega=0$, shows that 
the `pseudo' steady state persists on much longer time scales than the initial transient behaviour.
This may allow for the experimental observation of deviations from the steady state,
which here occur over $300$ periods up to the maximal propagation time.
To allow for such long propagation time in our calculation, we choose long leads with $L > 300 \times T = 6000$. The full system has $L_x+2L > 12500$ sites.

\begin{figure}
\centering
\includegraphics[width=0.5\textwidth]{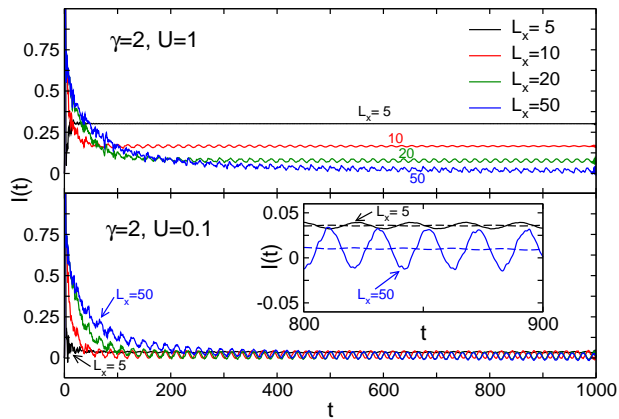}
\hspace*{2pc}
\begin{minipage}[b]{0.43\textwidth}
\caption{\label{disorder}
 (Colour online)
Current $I(t)$ at the contact surface of a disordered 3D device with $L_y=L_z=3$ and different $L_x$, for $\gamma=2$ and $k_F= \pi/4$.
A static voltage bias $U_L=U/2$, $U_R=-U/2$ is applied,
with $U=1$ (upper panel) and $U=0.1$ (lower panel).
The inset compares the $L_x=5$ to the $L_x=50$ system, and shows the time averaged current (dashed curve) in addition.
Note that the data are given for a single disorder configuration.
}
\end{minipage}
\end{figure}

In our second example in figure~\ref{disorder}, a constant voltage bias $U$ transports electrons 
through a disordered 3D device, with (static) random potentials with uniform probability distribution $V_\mathbf{m} \in [-\gamma, \gamma]$.
A steady state is approached only for the short system $L_x=5$ with large bias $U=1$.
Longer systems ($L_x \ge 10$) support localized states, whose presence causes oscillations in $I(t)$.
These become more pronounced for smaller bias $U=0.1$. 
The net current through the device is close to zero, since localized states do not contribute to transport.
Note that the average current for $L_x=5$ is much smaller for $U=0.1$ than for $U=1$, while for the $L_x=50$ system it is of the same magnitude.

In the given examples, up to $18450$ lattice sites are treated for device and leads.
Because of linear scaling in our method we must keep only that many elements in memory, and all calculations can be performed on standard desktop computers.
With quadratic scaling, we would have to store about $\approx 1.7  \times 10^8$ elements, requiring already $2.5$ GByte for complex numbers in double precision. 

In conclusion,
the numerical method discussed here allows for the calculation of time-dependent currents in large systems of non-interacting electrons, which are inaccessible to most other existing techniques.
The propagation of operators backwards in time is a tolerable disadvantage in such cases. 
Apart from these achievements, the development of advanced solution techniques for 
complicated non-equilibrium Green function equations of motion appears more promising for future progress,
especially with respect to the inclusion of dissipation or interaction.
How such developments are possible within the context of the Chebyshev Space method~\cite{AF08} and Sparse Polynomial Space approach~\cite{AF09} will be discussed elsewhere. Data obtained with the present method serve as a reference to validate these new techniques.

\ack
The authors acknowledge support by Deutsche Forschungsgemeinschaft
through SFB 652.

\section*{References}

\providecommand{\newblock}{}

\end{document}